\documentclass[sigconf]{acmart}
\usepackage{soul}
\usepackage{hyperref}
\usepackage[hyphenbreaks]{breakurl}

\copyrightyear{2025} \acmYear{2025} \setcopyright{acmlicensed}\acmConference[CHI '25]{CHI Conference on Human Factors in Computing Systems}{April 26-May 1, 2025}{Yokohama, Japan} \acmBooktitle{CHI Conference on Human Factors in Computing Systems (CHI '25), April 26-May 1, 2025, Yokohama, Japan} \acmDOI{10.1145/3706598.3713500} \acmISBN{979-8-4007-1394-1/25/04}

\begin{document}

\title{Beyond Automation: How UI/UX Designers Perceive AI as a Creative Partner in the Divergent Thinking Stages}

\author{Abidullah Khan}
\authornote{Both authors contributed equally to this research.}
\email{abid-ullah.khan@polymtl.ca}
\orcid{0000-0002-2131-9497}
\affiliation{%
  \institution{Polytechnique Montreal}
  \city{Montreal}
  \state{Quebec}
  \country{Canada}
}

\author{Atefeh Shokrizadeh}
\authornotemark[1]
\email{atefeh.shokrizadeh@polymtl.ca}
\orcid{0009-0004-0743-6875}
\affiliation{%
  \institution{Polytechnique Montreal}
  \city{Montreal}
  \state{Quebec}
  \country{Canada}
}

\author{Jinghui Cheng}
\email{jinghui.cheng@polymtl.ca}
\orcid{0000-0002-8474-5290}
\affiliation{%
  \institution{Polytechnique Montreal}
  \city{Montreal}
  \state{Quebec}
  \country{Canada}
}

\begin{abstract}
Divergent thinking activities, like research and ideation, are key drivers of innovation in UI/UX design. Existing research has explored AI's role in automating design tasks, but leaves a critical gap in understanding how AI specifically influences divergent thinking. To address this, we conducted interviews with 19 professional UI/UX designers, examining their use and perception of AI in these creative activities. We found that in this context, participants valued AI tools that offer greater control over ideation, facilitate collaboration, enhance efficiency to liberate creativity, and align with their visual habits. Our results indicated four key roles AI plays in supporting divergent thinking: aiding research, kick-starting creativity, generating design alternatives, and facilitating prototype exploration. Through this study, we provide insights into the evolving role of AI in the less-investigated area of divergent thinking in UI/UX design, offering recommendations for future AI tools that better support design innovation.
\end{abstract}



\maketitle 
\section{Introduction}
In today's rapidly evolving digital landscape, User Interface and User Experience (UI/UX) design play a pivotal role in shaping user interactions, significantly influencing product success and overall user satisfaction~\cite{hassenzahl_user_2008, Hartson2019}.
A critical component of the UI/UX design process is divergent thinking, which involves exploring a wide range of ideas and solutions to drive innovation and creativity~\cite{frich_how_2021}. Different from convergent thinking, which prescribes operations to help achieve ``a single correct solution''~\cite{Finke1992}, divergent thinking ``allows one to explore in different directions from the initial problem state, in order to discover many possible ideas and idea combinations that may serve as solutions''~\cite{Finke1992}. While both aspects are important in UI/UX design~\cite{designcouncilDoubleDiamond}, divergent thinking allows designers to break free from conventional patterns and explore novel approaches to problem-solving, a process vital for tackling the current friction between rapidly advancing technology and limited design innovation~\cite{Dove2017, Yang2018}.

Over recent years, the integration of artificial intelligence (AI) tools into various stages of the UI/UX design process has become increasingly prevalent. These AI tools offer new possibilities for ideation, prototyping, and refinement~\cite{sermuga_pandian_metamorph_2021, latipova_artificial_2019, takaffoli_generative_2024, Mozaffari2022, Sobolevsky2023}. Recent research has also significantly advanced our understanding of AI's role in the design process. For instance, \citet{zhou_how_2023} demonstrated AI's potential to enhance both the depth and breadth of design solutions through strategic integration points in the design process. Similarly, \citet{chandrasekera_can_2024} illustrated how generative AI can boost creativity and reduce cognitive load, particularly within design education, underscoring AI's role as a co-creator in producing diverse and innovative design solutions. With the recent advancements of generative AI, many commercial tools also emerged to aid the UI/UX design process. For example, tools like ChatGPT streamline data collection and synthesis, while MidJourney, DALL-E, and Uizard assist in generating visual assets and prototypes. Additionally, Figma's AI plugins like Automator\footnote{https://www.figma.com/community/plugin/1005114571859948695/automator} improve workflows by automating repetitive tasks. Tools like Colormind\footnote{http://colormind.io/} also offer creative inspiration through palette generation.

Despite the significant advancements in AI-assisted design, there remains a critical gap in understanding how professional UI/UX designers perceive AI in supporting divergent thinking, as well as their needs and desires in AI tools to help them create more innovative products. Moreover, while previous studies have explored AI's contributions to design iteration~\cite{zhou_how_2023} and early-stage creativity~\cite{chandrasekera_can_2024}, little attention has been paid to how AI tools are currently utilized during the divergent thinking phases in the actual design practice. To address this gap, our study deliberately focuses on this aspect and investigates three key research questions: 
\begin{itemize}
    \item 
\textbf{RQ1}: What are the current practices of UI/UX designers in the divergent thinking process?
\item
\textbf{RQ2}: How do designers use AI tools to support divergent thinking in the UI/UX design process?
\item
\textbf{RQ3}: What are designers' perceived values and desires in using AI to enhance divergent thinking in their UI/UX design process?
\end{itemize}

To explore these questions, we conducted in-depth interviews with 19 experienced UI/UX designers with diverse backgrounds and expertise. Analysis revealed that designers emphasized research, inspiration, and prototype exploration as crucial components of their divergent thinking process. They often drew inspiration from a variety of sources and employed iterative approaches to explore and refine their ideas. AI tools, while increasingly used to accelerate tasks such as research data collection and synthesis, idea generation, and prototype creation, are seen as assistants that support, rather than replace, the divergent thinking process. Participants expressed a strong desire for AI tools that offer more efficient collaboration features, customized support, and visual interaction, while ensuring that designers retain full control over the creative process. Our study contributes to the understanding of practitioners' unique perspectives for using AI tools in divergent thinking, including the need to balance efficiency with creativity, the uncommon lack of concerns about copyright and ownership, and the blurred distinctions between low-fidelity and high-fidelity prototyping approaches. Our results also indicated four key roles AI plays in supporting divergent thinking: to aid research, kick-start creativity, generate design alternatives, and alter prototype fidelity for feedback. For each role, we contribute opportunities for future AI tools to facilitate related activities. Overall, our study offers valuable insights into how AI tools can be better integrated into the divergent thinking process to reshape creative practices in UI/UX design.

\section{Related Work}
\subsection{Divergent Thinking in UI/UX Design and Creativity}

The UI/UX design process focuses on prioritizing a user-centered approach that iteratively refines solutions based on user feedback and testing. This process typically includes several key stages: research, ideation, prototyping, and testing, each of which is crucial for ensuring that the final product effectively meets user needs~\cite{hunt_what_2023}. Among the frameworks guiding this process, the Double Diamond model stands out as a structured approach to capture both divergent and convergent thinking processes~\cite{designcouncilDoubleDiamond, banbury_using_2021}. Developed by the UK Design Council, the Double Diamond model divides the design process into four phases: Discover, Define, Develop, and Deliver~\cite{designcouncilDoubleDiamond, nessler_how_2018}; the Discover and Develop phases are considered to be focused on divergent thinking, while the other two convergent thinking. Overall, this model emphasizes the importance of divergent thinking in the initial stages, where a broad range of possibilities is explored before converging on a defined problem~\cite{frich_how_2021, hennigan_optimizing_2019}. 

Several researchers have investigated the role of divergent thinking in design, focusing on its cognitive foundations and practical implications. For instance, \citet{xie_cognitive_2023} conducted a detailed study on the cognitive processes that drive divergent thinking, particularly contrasting the approaches of expert and novice designers. Their work shows that expert designers employ structured cognitive strategies, leading to more innovative outcomes, whereas novices often struggle with these strategies, resulting in less creative solutions. This research underscores the importance of cognitive approaches in enhancing creativity during the design process. \citet{goldschmidt_linkographic_2016} expanded on this understanding by examining the interplay between divergent and convergent thinking in design. Through a linkographic analysis, a method developed by Goldschmidt to map out and analyze the dynamic connections and shifts in design ideas during brainstorming, it is demonstrated that changes between these cognitive modes occur so frequently during ideation that they can be considered concurrent rather than sequential. This insight challenges the traditional view presented by the Double Diamond model and suggests that design thinking is a more fluid and interconnected process, which could influence how design education and practice approach the training of future designers. \citet{cromwell_how_2024} explored the impact of constraints on creativity, particularly within the framework of divergent thinking. His research suggests that creativity can be enhanced by a balanced combination of constraints, such as those related to problems and resources. This balance is crucial in fostering motivation and encouraging creative exploration, aligning with the Double Diamond model's emphasis on using both divergent and convergent thinking to manage constraints effectively. Furthermore, the 'order effect,' which describes the evolution of ideas during the ideation process, has been extensively investigated. According to \citet{kaya_impact_2019}, ideas become more original as they go through the ideation process. This is due to the activation of more complex cognitive processes, such as imagination, rather than relying exclusively on memory. This finding emphasizes the necessity of encouraging designers to think beyond their original ideas and explore deeper creative possibilities. Several tools have been created to help designers overcome difficulties like design fixation ~\cite{zuo_wikilink_2022,huang_mitigating_2021,chen_asknaturenet_2024}. For example, \citet{chen_asknaturenet_2024} developed AskNatureNet, a tool that supports divergent thinking in design ideation using bio-inspired design knowledge. This tool incorporates biological knowledge into a semantic network, allowing designers to draw comparisons from nature, overcoming design fixation and widening the range of potential solutions.

In sum, the literature highlights the importance of supporting divergent thinking in UI/UX design. Our research builds on this by exploring how AI can be integrated into the divergent thinking stages to support these cognitive strategies, aligning with and advancing the creative processes identified in existing studies.

\subsection{Using AI in UI/UX Design}
To date, several studies have investigated the integration of Artificial Intelligence (AI) into the UI/UX design process, particularly in how it enhances creativity, automates design tasks, and addresses specific design challenges. These studies reflect a broader trend in which AI is increasingly seen as beneficial in the creative problem-solving aspects of the Design Thinking process~\cite{saeidnia_integrating_2024, poleac_design_2024,liao_framework_2020,padmasiri_ai-driven_2023}. For instance, \citet{padmasiri_ai-driven_2023} demonstrated how AI tools enhance stages such as understanding user needs, rapid prototyping, and iterative testing. Building on this, \citet{bertao_artificial_2022} discovered that there is a growing recognition of AI's potential to increase process efficiency, particularly when used as a virtual assistant in UI/UX design. Expanding on AI's role as a creative partner, \citet{chiou_designing_2023} demonstrated that AI tools enable a broader exploration of design possibilities. Unlike \citet{padmasiri_ai-driven_2023}, who focused on prototyping and testing, they prioritized the ideation phase, in which generative AI models provide unique perspectives and develop alternate design ideas. Similarly, \citet{tholander_design_2023} found that generative AI improves the ideation process by providing new inputs that might not be immediately obvious to human designers.

Various AI tools have been investigated to improve numerous areas of the design process. For instance, \citet{york_evaluating_2023} evaluated the application of ChatGPT in UX design and web development education, noting that AI has been utilized in tasks such as brainstorming, design, and coding. Their results revealed that while ChatGPT performs well in brainstorming and ideation tasks, its effectiveness diminishes in more complex areas such as detailed design work and coding, where the outputs often require significant human oversight and correction. Similarly, \citet{sermuga_pandian_metamorph_2021} introduced MetaMorph, an AI tool designed to convert low-fidelity sketches into higher-fidelity versions; using this tool, designers reported above-average satisfaction due to the time and effort saved. These studies collectively suggest that AI's ability to enhance efficiency is widely recognized, though the practical application of these tools may vary across different stages of the design process. 

Despite these acknowledged benefits, the integration of AI in a UI/UX design process introduces significant ethical concerns. \citet{chaudhry_concerns_2024} examined these implications, noting that while designers value the efficiency AI brings, there are ongoing concerns about its potential to disrupt creative processes. Complementing this perspective, \citet{inie_designing_2023} emphasized the importance of participatory AI design to ensure that these tools not only adhere to ethical standards but also meet the specific needs of designers. Similarly, \citet{li_user_2024} highlighted potential drawbacks, revealing that while GenAI tools boost creativity and efficiency, experienced designers have significant concerns about potential skill degradation, job displacement, and ethical issues like copyright and ownership. This contrast underscores the ongoing debate about AI's impact on the long-term development of design skills. Furthermore, building on the concerns about AI's impact on team dynamics, \citet{wang_exploring_2024} explored the impact of AI tools for content generation on social interactions within UX teams. They discovered that, while these tools improved communication and collaboration, they also created concerns about the reliability and quality of AI-generated outputs, complicating the ethical perspective of AI in design.

There is also a notable gap between AI research and its practical application in the design industry. \citet{lu_bridging_2022} highlighted that many AI-enabled design tools, despite their theoretical promise, often fall short in addressing the specific needs of UX professionals, particularly in tasks that require design thinking rather than just generating graphical outputs. Expanding on this issue, \citet{uusitalo_clay_2024} examined how generative AI tools are perceived by UX and industrial designers, emphasizing the need to enhance designers' sense of control when using these tools. This aligns with \citet{bertao_artificial_2022} insight that as AI tools become more widespread, there will be a shift from viewing AI merely as a facilitator to recognizing it as a collaborative partner in the design process. Systematic reviews, such as the one conducted by \citet{shi_understanding_2023}, have categorized existing AI design support systems (AI-DSS) and identified key themes in how AI is used to assist in design. Their review highlights AI's potential to enhance human creativity, recommending that future research should prioritize developing AI tools that are more explainable, ethical, and adaptable to various design contexts. Similarly, building on the importance of adaptability, \citet{rodriguez_prado_real-time_2024} and \citet{wang_exploring_2024} emphasized the necessity for AI tools that not only support the technical aspects of design but also improve team collaboration. These studies collectively highlight the critical need for AI tools that are not only technically proficient but also aligned with the ethical and social dynamics of the design process.

While the existing literature extensively explores the various roles AI plays in the design process, there remains a critical need for research that bridges the gap between AI's theoretical promise to support the divergent thinking process of UI/UX design and its limitations during practical applications in this process. Our study aims to address this gap by focusing on understanding designers' experiences and perceptions regarding using AI tools in divergent thinking and suggesting directions for future tool design to better support this practice. By doing so, we hope to contribute to a deeper understanding of AI's role as both a creative partner and an efficiency-enhancing tool for supporting divergent thinking in the UI/UX design process.
\section{Methods}
The study is approved by the Research Ethics Board of the involved institutions. Below, we describe our participants, the interview approach, and data analysis methods.

\subsection{Participants} 
To recruit participants for this study, we targeted professional UI/UX designers with a minimum of one year of experience and at least one completed design project. We used multiple channels to recruit our participants, including advertising our study on professional networks like LinkedIn and social media platforms such as Facebook, as well as direct outreach to professionals within our networks. Ultimately, we recruited 19 participants (9 females, 10 males), ranging in age from their 20s to 40s. Each participant received an incentive of \$30 CAD. To maintain confidentiality, unique identifiers (P1 to P19) were assigned to participants and used throughout the study. Table 1 details the characteristics of the participants.

\begin{table*}[t]
\centering
\small
\caption{Summary of Participants' Characteristics}
\label{tab:participant-info}
\resizebox{\textwidth}{!}{%
\begin{tabular}{cclcll}
\toprule
\textbf{ID} & \textbf{Gender} & \textbf{Role/Profession} & \textbf{Yrs of Exp.} & \textbf{Key Projects} & \textbf{Education/Background} \\ \midrule
P1  & Female & UI/UX Designer & 5 years & Platforms for finance and agriculture & Bachelor's in Architecture \\ 
P2  & Male   & Product Design Lead & 10 years & Games, web apps, and cybersecurity apps & Master's in UI/UX design \\ 
P3  & Male   & UI/UX Designer & 8 years & Financial technology products & College diploma in Computer Science \\ 
P4  & Female & UI/UX Designer & 4 years & Wearable apps, e-commerce platforms & Bachelor's in Industrial Design \\ 
P5  & Female & UX Designer & 6 years & Event-booking apps, clothing websites & Master's in Interior Design \\ 
P6  & Female & UX Designer & 2 years & Internal tools for engineering teams & Master's of Architecture\\ 
P7  & Male   & Product Designer & 2 years & Mutual funds investment platforms & Master's in Interactive Media \\ 
P8  & Female & Graphic/UX/UI Designer & 10  years & E-commerce platforms & College diploma in Graphic Design \\ 
P9  & Male   & UX Designer & 8 years & B2B tools, document comparison tools & College diploma in Animation \\ 
P10 & Female & Interaction Design Student & 2 years & Small-scale projects, academic projects & Bachelor's in Interaction Design \\ 
P11 & Male   & UX Designer/Teacher & 15 years & Telecom enterprise apps, web apps & Bachelor's in UX Design \\ 
P12 & Female & UX Junior Consultant & 2 years & Streaming platforms & Master's in UX Design \\ 
P13 & Male   & Product Designer & 4 years & Financial technology products & Bachelor's in Business and Marketing \\ 
P14 & Female & UI/UX Designer & 4 years & Various iOS apps and websites & Master's in Cognitive Neuroscience \\ 
P15 & Female & UI/UX Designer & 2 years & Small-scale projects, academic projects & Master's in UX Design \\ 
P16 & Male   & UI/UX Designer & 2 years & Educational apps & Bachelor's in Mechanical Engineering \\ 
P17 & Male   & Graphic Designer & 2 years & Web applications and dashboards & Master's in Management Science \\ 
P18 & Male   & UI/UX Designer & 6 years & Crypto exchange platforms & Bachelor's in Computer Science \\ 
P19 & Male   & UI/UX Designer & 2 years & Various mobile and web apps & Bachelor's in Management Science \\ 
\bottomrule
\end{tabular}%
}
\end{table*}

\subsection{Data Collection}
We conducted semi-structured interviews with the participants. The interview questions were designed to gain a deep understanding of (1) how participants currently perform divergent thinking in design, including how they started their design process, handled constraints, created and managed artifacts, and engaged in collaboration, and (2) how they used and perceived AI tools in these activities. Before conducting the main study, we performed a pilot interview with a professional UI/UX designer to test the interview questions. The pilot interview allowed us to refine the questions, improve clarity, and ensure that the interview flow supported an in-depth exploration of the study's objectives. Feedback from the pilot interview informed adjustments to the final interview guide. The specific questions in the final interviews were divided into three main sections. The first section aimed to gather demographic information about the participants, including their professional experience, educational background, and the types of UI/UX projects they had worked on. The second section explored how participants currently approached divergent thinking in their design process, including the general activities they carried out to support divergent thinking, the impact of collaboration in these activities, the strategies they used to handle various constraints, the physical and digital design artifacts they created and managed in those activities, and the tools they used. The third section aimed to explore how AI was integrated into participants' divergent thinking processes, the advantages and limitations they observed, and their overall perceptions of AI's role in UI/UX design. 

All participants were sent a consent form via email prior to their interview. The consent form outlined the purpose of the study, participant rights, and data management protocols. Participants were required to review, sign, and return the consent form via email before the interview. Only participants who provided signed consent were included in the study. The first two authors conducted all interviews jointly to ensure a comprehensive capture of participants' insights and to avoid missing any important information. All interviews were conducted remotely using Microsoft Teams and Zoom; each session lasted 45 to 60 minutes. With the participants' consent, both audio and video of the interviews were recorded.

\subsection{Data Analysis} 
Each interview was initially transcribed using the Microsoft Word Transcribe tool. We carefully reviewed and manually corrected errors in the automated transcription to ensure accuracy. We then conducted thematic analysis~\cite{Vaismoradi2013, aronson_pragmatic_1995} on the transcribed interview data to answer our research questions. The analysis focused on identifying key themes and categories about (1) participants' current practice in divergent thinking, (2) their current use of AI to support divergent thinking, and (3) their values and desires for AI tools. The coding process was carried out using ATLAS.ti\footnote{https://atlasti.com}, a qualitative analysis software that enabled us to efficiently organize, categorize, and retrieve data throughout the analysis.

The coding process was essentially iterative and inductive. It began with each author independently familiarizing themselves with the data by thoroughly reading the interview transcripts. Following this, we conducted an initial round of open coding using ATLAS.ti. Each author applied descriptive codes to text segments to capture the main ideas discussed by the participants. This approach was carried out iteratively by each individual author. After completing the first round of coding, we moved into a collaborative phase where all authors participated in meetings to review, discuss, and refine the codes assigned during the initial stage. The objective of these meetings was to ensure that the codes accurately captured meaningful insights to answer our research questions and address the overall research objective. During these discussions, we grouped codes into broader, more encompassing categories. This process was supported by an affinity diagram created in Miro, to visually map out the relationships and connections among the identified codes and categories. In cases of disagreements, we conducted in-depth discussions where each author presented their interpretation of the text segments and all authors provided inputs to attempt to reach consensus through open communication.

As we progressed with the analysis, we continued employing an inductive approach to identify and extract themes from the codes and categories. When new insights emerged, we revisited our coding on previously coded transcripts to make adjustments. We repeated this iterative process until no new themes or categories emerged. We conducted several collaborative meetings to review the affinity diagram to ensure that it accurately captured an accurate reflection on the participants' practices and experiences. In these discussions, we further revisited and adjusted the codes, categories, and themes as needed. The final set of categories and themes generated from this process were reported as results below.

\section{Results}
\subsection{RQ1: Designers' Current Practice in Divergent Thinking}

\subsubsection{Research and Discovery} (The number of participants mentioned this theme is $N=9$.)
This is the foundational activity in which UI/UX designers explore and gather information essential for creative ideation and problem-solving in design projects. Several participants emphasized user research as a crucial component of the divergent thinking process. For instance, P5 highlighted the importance of understanding user needs, stating, ``\textit{One thing that I missed out on before is actually having a clear set of user personas. So, there you have a clear understanding of what the frustrations of each user are.}'' Moreover, defining project scope and objectives emerged as another critical aspect of this activity in the divergent thinking process. P2, for example, noted that the main goal of initial stakeholder engagement was ``\textit{to learn the objectives, mission, and the problem they want to address to create a product for it.}'' 

\subsubsection{Inspiration and Ideation} ($N=14$)
For UI/UX designers, creativity in divergent thinking requires stepping away from traditional approaches to discover novel methodologies that push design boundaries. Our findings indicate that designers frequently use platforms like Behance and Dribbble for inspiration. For example, P14 stated, ``\textit{Sometimes I use other sites like Dribbble, but mostly Behance. The reason is that Behance often shares the design process, while Dribbble usually focuses on the final product.}'' Similarly, designers often draw inspiration from existing designs to refine and improve their own designs. As P17 noted, ``\textit{We see if something similar has been done in the past by someone and then we just take inspiration from that.}'' Related, market research emerged as another important factor of divergent thinking that offers insights into industry trends and understanding competitive dynamics. For example, P3 listed a series of questions he asked: ``\textit{Usually, I start by understanding what the project is all about, like... Is this something that's first to market? Does it already exist? Are there benchmarks we can compare it to? ... All those details to really get a sense of what we're dealing with.}'' 

Collaboration is another important factor that enhances the ideation process, enabling the exploration of diverse ideas, which is essential for divergent thinking. Participants agreed that such communication is essential for generating creative solutions. For example, P12 stated, ``\textit{We always try to keep an open space, making sure everyone is comfortable sharing their ideas and leaving space for anyone to talk. We bounce off each other, like `Hey, what do you think?' or `Could you maybe note this down?' ... Being open with each other is the biggest part of our collaboration.}'' In remote settings, digital tools can play an important role in facilitating collaboration during brainstorming sessions. As P2 mentioned, ``\textit{I use Miro boards and FigJam for brainstorming sessions with both designers and non-designers.}'' 

\subsubsection{Exploring Design Options: Lo-fi/Hi-fi Prototyping} ($N=15$)
Exploring design options through prototyping is often considered the final, yet essential phase of divergent thinking by our participants.
Participants generally emphasized the importance of starting with traditional methods, such as pen-and-paper sketches, to visualize initial ideas, considering digital tools as distracting or limiting in the initial exploration. For example, P8 stated, ``\textit{The first step in anything creative is to let things flow naturally. It's like free writing. Ideas come out quickly, and it's hard to capture them directly on the computer because of the tool interface.}'' Similarly, P4 highlighted the limitations of digital tools in exploring design ideas, stating, ``\textit{I can easily change the design with my pen or pencil, which makes it super helpful for the first test and for exploring ideas. Then I move on to wireframing, which doesn't have color, but it takes time [to make]... That's why I prefer to work on paper first before moving to a digital platform.}''

Once initial ideas are visualized and iterated upon, gathering feedback from users and other stakeholders becomes essential for further exploration. Participants described beginning their design process with paper artifacts, which they enriched through iterative feedback. For example, P19 shared, ``\textit{I draw wireframes on paper at the initial stage of my design process... Because if we draw 7 to 8 wireframes, it is very helpful for evaluating the design or getting feedback. I share the paper sketches or wireframes with other designers, and they provide their comments and feedback at this early stage of the design process.} Digital tools were sometimes adopted to help with feedback gathering and collaboration. For example, P2 emphasized this by saying, ``\textit{We rely heavily on the comments tool in Figma. You can tag someone and assign tasks, which is really effective.}''

\subsection{RQ2: The Current Use of AI in Divergent Thinking}
\subsubsection{Pre-Design UX Activities} ($N=6$)
UX practitioners have leveraged AI tools as assistants in pre-design activities, such as research, data analysis, and the creation of pre-design artifacts. These usages are aimed to support the unique demands of UX divergent thinking. For instance, some participants used ChatGPT to identify competitors. As P3 mentioned, ``\textit{I've already used it to benchmark. So, like find me all the best examples that have this experience as a checkout or something along those lines.}'' This method allows them to complete time-consuming tasks more quickly and efficiently, freeing up time and energy that can be redirected towards more creative activities. About this, P5 noted, ``\textit{Just going back three to four years ago, to get an insight or to have a normal secondary research, you would have to go to 10, 15 websites to get one piece of information. But now with AI and tools such as ChatGPT, it's so easy to prompt something and get data that would have probably taken two days to collect. ... It's quick, efficient, and helps me focus on other aspects of the project. It not only makes me more productive but also enhances the quality of my work.}'' Additionally, AI is used to analyze data collected during user research to help designers gather insights to inform the creative process, as P6 explained, ``\textit{I use ChatGPT just to like brush up on my research, mostly when I'm in the research part of the project. For example, recently I was doing user research, and what I did was use ChatGPT to summarize some of the findings...}'' They even sought AI assistance in generating pre-design artifacts so that they could quickly explore user types and usage scenarios, as well as generate artifacts for activities such as brainstorming. For example, P6 noted its use in creating user personas based on research data and P10 used AI to create mind maps: ``\textit{I just wrote my prompts, and it gave me an endless number of questions related to that problem... So you know you have a problem statement, then you have to start ideating about it, iterating around it.}'' P11 also used AI to create artifacts for facilitating collaboration during ideation: ``\textit{I used AI in Figjam to generate worksheets for ideation sessions -- when I bring other people in to collaborate, ask for their ideas, and vote on them. Figjam can create the whole framework by typing a simple request, and it generates a visual canvas to be filled out.}''

\subsubsection{General Inspiration on Design Solutions} ($N=5$)
Participants mentioned that when they lacked a concrete idea or were not sure where to start, they turned to AI tools for brainstorming and ideation support. For example, P19 discussed using various AI tools for inspiration: ``\textit{I use Freepik to get images and vector illustrations, and Framer AI to gather business-related content. When designing digital marketing websites, I often turn to Awwards.com for inspiration. Additionally, I use ChatGPT to generate content for web pages, which I then integrate into my designs.}'' However, many participants, such as P13, expressed that ``\textit{AI gives us some ideation. But ... it's not a final solution for me.}'' This reflected a common approach among participants during divergent thinking: using AI-generated ideas as a foundation for further exploration. They understood that while AI can produce a wide range of ideas, the human touch is crucial in refining these into practical design solutions. Furthermore, participants sometimes used AI after they had initial ideas to help organize and expand on them. As P4 explained, ``\textit{I use AI tools for managing some of my ideas and my thinking. For example, for categorizing ideas, for getting some new suggestions, and for receiving suggestions for other methods that are suitable for particular problems.}'' In this way, AI tools assist not only in generating new ideas but also in structuring and enhancing existing ones. 

\subsubsection{Ideating on Specific Design Aspects}($N=7$) 
Participants also used AI tools to gather rapid feedback and diverse options on specific design aspects. For example, some participants used AI to generate ideas on what elements to include on a UI screen. On this, P15 shared, ``\textit{It helps me mostly when I'm working on wireframes. For example, while building a new website, instead of just researching or creating a mood board, I asked simple questions, like, what elements or design features should I use? It's another source of exploration.}'' Similarly, designers turn to AI for layout ideas. As P7 mentioned, ``\textit{I was trying to just see how it is. To get the layout right, get the structure and so on, it was good.}'' Another area where AI is used is in selecting a color palette, as P10 explained, ``\textit{Just before we're moving towards the prototype stage, I usually go to MidJourney to get inspiration. It helps me explore ideas for illustrations or find color palettes that could be used in the design.}'' Finally, participants also utilized AI for ideating textual and visual content. For example, P10 described using AI for visual inspiration: ``\textit{I used MidJourney a lot for getting inspirations and just randomly trying to type in the prompts according to the need -- MidJourney was born for getting the visuals.}'' By generating ideas across various design aspects, AI tools enabled designers to compare different approaches quickly, which served as a crucial factor in stimulating divergent thinking.

\subsubsection{Prototyping and Refinement}($N=7$)
Our participants reported using AI tools when exploring design ideas through creating prototypes, often at the end of the divergent thinking process. Some have experimented with AI tools like UIzard to build complete prototypes from scratch, customizing them according to their needs. As P11 noted, ``\textit{I've been a little bit curious about design tools starting to come out like UIzard. It's able to create visual, clickable prototypes from a text prompt, which saves me hours or even days of work. It's very, very good.}'' Participants considered these AI-powered tools as valuable because they provide a solid starting point, eliminating the need to begin from a blank page and significantly speeding up the design process. On the other hand, some designers take a different approach by creating the initial prototypes themselves and then using AI tools to explore more options and refine their existing work. P15 explained, ``\textit{You write some prompt about, like, getting a button for me, and I put this color palette, etc. You can input even the hex code to change the color that you want.}'' This demonstrates AI's ability to facilitate quick iterations, enabling designers to experiment with various elements and play with design ideas without the need for time-consuming manual adjustments.

\subsubsection{Creating Textual Content on Prototypes} ($N=6$)
Participants mentioned that they have increasingly relied on AI tools to generate various types of textual content, such as text for tooltips, pop-ups, titles, and more. P15 highlighted an interesting effect of using AI for this purpose in early prototyping, stating, ``\textit{Usually before that, when you have something, you just use lorem ipsum, like random text. With AI, you can reduce that amount of back and forth with a copywriter. It's not a final, but you have something more realistic, closer to the realistic design, so that you can work on it. The framing, how long the text can go, a lot of things that you have to take into account.}'' This reflects how AI-generated content allows designers to work with text that is more contextually appropriate so that they can incorporate this factor when considering other design elements before consulting other stakeholders. P8 added to this, emphasizing the usefulness of having realistic text to gather meaningful feedback: ``\textit{Yeah, I might use ChatGPT to get, you know, a little description that I don't want to have to write out myself as a base to show the client what direction to go in. And usually from there, they'll fiddle around with that kind of text.}'' 

In smaller companies or for freelancers where a copywriter may not be available, however, UI/UX designers often take on the responsibility of creating the textual content themselves. This often involves more divergent thinking, as the designers need to ideate on the tone and context and align them with the brand. Participants mentioned that they used AI tools like ChatGPT to assist in this creative process. As P9 explained, ``\textit{It's not necessarily that I grab whatever it spits out, it's more like I'm messaging back and forth with it to get it to... so that it's in the tone of the brand as well, and then it's communicating what it should depending on its context.}'' P14 also highlighted the importance of AI tools like ChatGPT for creating concise and effective content, stating, ``\textit{I always use ChatGPT for the writing parts, because I always try and do my best to be straight to the point in the writing.}'' This iterative interaction with AI tools fosters a more exploratory approach, enabling designers to experiment with different textual variations until they find the most suitable content for their design.

\subsubsection{Creating Visual Content on Prototypes}($N=5$)
When participants wanted to incorporate visual content, such as illustrations, into their prototypes, they often turned to AI tools like DALL-E and MidJourney. These tools allowed them to transform their ideas into visual representations, a process that was much more challenging before the advent of AI. P5 highlighted these difficulties: ``\textit{I remember before, where I used to do it, say I need women carrying a bag in an animated form but smiling. Like I have such particular things, but when I'm going to put it on Google, I'm not going to get exactly that. And even if I do, it's going to be similar to many other people who have already used it in their portfolios or in their designs or UI screens.}''
With AI, designers can now generate more tailored images that meet their exact specifications. In some cases, they used the generated visuals directly in their prototypes, as P5 mentioned, ``\textit{If I need something and I want people to actually believe that this is an actual image, I can still do it [with AI].}'' 
However, in many cases, AI-generated visuals were not perfect, so designers used these tools more for inspiration rather than as final solutions. For example, P4 explained, ``\textit{It was helpful for me for inspiration and then designing by myself... -- just for that. If I want to [directly] use the image that it suggested, it has a lot of mistakes that we can find.}'' In this way, AI tools support the exploratory phase of design by providing a starting point for ideation, allowing designers to refine and build upon the initial concepts generated by AI.

\subsubsection{Not Using AI in Professional Practice} ($N=8$)
While many UI/UX designers are adopting AI tools in their divergent thinking activities, some of our participants chose not to use them in their professional practice for several reasons. For some, AI simply is not necessary for their current roles. For example, P6 explained, ``\textit{Whenever I see like in LinkedIn or other platforms, when they introduce them... I go to the tool itself and see what they like, what is the purpose. But I haven't used any of them really, maybe because in my current job I haven't needed them so far.}'' Similarly, P16 also stated, ``\textit{I have used AI tools in the past, but in our current UI/UX work, I haven't found the need to use any.}''
Concerns about the quality of AI-generated outputs also deter many designers, with P18 stating, ``\textit{I didn't use them because they are so blurry and smudgy, there's no detail in it, and realism is not there.}'' P17 also explained: ``\textit{A critical aspect is accuracy, especially for specialized applications in specific companies. A generalized language model often isn't suitable for these unique needs, which is why many companies, including mine, are developing their own customized models.}'' Additionally, a lack of awareness and familiarity with AI tools contributes to hesitation. For instance, P7 mentioned, ``\textit{It's something that I need to research about more and learn about most of mine. So yeah, I think that's one of the main reasons I'm still holding back from using it at work and even me personally as of now.}'' Company restrictions, sometimes related to data safety and privacy, also limited access, as P6 noted, ``\textit{they have so many guidelines for safety and security, which affects my curiosity about AI as well.}''
Finally, financial constraints make accessing advanced AI tools difficult, especially for freelancers. P5 pointed out, ``\textit{When it comes to image generation, it's a different story. You have to have a certain plan to be able to access that.}'' These factors collectively influence why some UI/UX designers opt not to incorporate AI into their design processes.

\subsection{RQ3: Values and Desires for AI in Divergent Thinking}
\subsubsection{Designer in Control} ($N=10$) Maintaining creative authority is essential for designers when integrating AI into the design process, especially in the context of divergent thinking. While AI can offer valuable assistance by generating diverse ideas, refining designs, or automating repetitive tasks, participants insisted that the designer must remain the primary driver of the creative process. They expressed a strong desire for AI tools that enhance their creativity without diminishing their role as decision-makers. For example, P5 highlighted, ``\textit{You have to be very mindful when you're using AI. According to me, I cannot rely on it 100\%... It's just like how AI can never replace real-time artists... As designers, I think we'll have to be empathetic towards that as well.}'' Similarly, P6 expressed a preference for AI as an assistant rather than a replacement, noting, ``\textit{I don't want to think of AI as tools to take over our creativity ... So yeah, I like to look at AI as an assistant to brush up your work.}''

One important aspect allowing designers to stay in control is the need for AI tools to generate a wide range of distinct and meaningful design outcomes for designers to examine, choose, alter, and combine. For example, P10 mentioned, ``\textit{Usually, there's a generic set of designs which the AI iterates a little bit and presents a new idea. If it helps in creating differentiable iterations, it will be more helpful to get more inspiration or a base for our design.}'' Expanding on this, P2 suggested, ``\textit{It could give me three or four content options generated by AI while keeping previous versions available for reference.}'' Additionally, P18 highlighted the importance of AI in assisting with the most challenging aspects of design, saying, ``\textit{The most difficult part in UI/UX design is making a hero section [the web UI area immediately below the navigation menu] look incredible. There should be an AI tool that provides variations and layouts for inspiration.}''

To ensure designers have control over the creative process, participants also highlighted the importance of AI tools that can allow designers to specify and manage design constraints like brand guidelines, color schemes, and industry-specific requirements. P5, for instance, suggested, ``\textit{It could ask for color codes that you're planning on using maybe, and then it could probably get a gist of the idea of the brand it is.}'' Expanding on this, P3 stated, ``\textit{If there was a way to consider different brands, colors, and versions, and say `create this with a different brand' or `take these screens and make them mobile-friendly,' that would be great.}'' P10 further emphasized the need for AI to tailor designs to industry-specific requirements, observing, ``\textit{The medical industry should have different visuals, color tones, and features to give a more personalized feel.}''

Moreover, participants emphasized the desire to be able to refine and adjust AI-generated designs to ensure that the final design aligns with their requirements. For example, P1 stated, ``\textit{As I mentioned before, I can specify what I need, but I also want to be able to edit it, like changing a specific thing or creating the next page in a certain way.}'' Similarly, P13 expressed a desire for more targeted control over individual design elements, saying, ``\textit{I want to select the illustration and just say that I want a realistic illustration. But when I use the keyboard to specify this to the AI tool, the next design it gives me changes everything on the screen -- the title, the description, the button color, everything.}'' Building on this, P1 shared similar insight stating, \textit{If we could define section by section or page by page what we want to change, like modifying something on the right side, it would be really helpful.}'' This indicates the importance of AI tools providing designers with the ability to make specific adjustments without losing control over other elements of the design.

\subsubsection{Supporting Collaborative Data Analysis and Artifact Management} ($N=9$)
Participants highlighted the transformative potential of AI tools in data analysis, particularly in integrating with existing databases, providing real-time insights, and enhancing understanding of data. For instance, P11 envisioned a future design process: ``\textit{Ideally, I want a tool where I can start a new project by asking what artifacts we have in our database, such as user research sessions or brand guidelines. Then, using this information, it can design the attempt page app accordingly.}'' Similarly, P11 mentioned the importance of AI in accessing and analyzing private data repositories, noting, ``\textit{It would be nice if an AI system could analyze a corporate database full of research and provide summaries or key points.}'' Additionally, P17 highlighted the need for more interactivity in AI-powered data analysis tools, stating, ``\textit{Generative AI tools could be more interactive, like Tableau or Power BI. It would be helpful if these tools could generate dashboards automatically based on prompts.}'' These capabilities would enable design teams to quickly draw on relevant information, streamlining collaboration and facilitating more efficient problem-understanding and problem-solving.

Participants also expressed a strong desire for AI tools capable of analyzing and interpreting real-time information during design sessions. As P12 suggested, ``\textit{Maybe if you had those notes that were generated and maybe that said, oh, like two of these participants said these key insights, like you should maybe continue talking about that or ask them questions about that. That would be kind of cool, kinda like an assistant.}'' Such tools could significantly enhance collaborative processes by guiding more efficient discussions and facilitating insights gathering. Additionally, AI's potential to participate in conversations was highlighted as a valuable feature, with P15 mentioning, ``\textit{It would be nicer if the AI can jump into the conversation and kind of interpret the underlying emotions and touch on them.}''

With the large number of artifacts created during the divergent thinking phases, organizing and documenting the various stages and revisions of a design project is another area where participants saw the potential for AI tools. For example, P12 said, ``\textit{When comparing different designs against each other, maybe the AI could help me identify what kind of structure is used, if there's a similarity between them, etc.}'' P2 also mentioned, ``\textit{It would be great if Figma had a feature where... there would be a versioning system to track different versions of the content. This way, you could revert to previous versions if necessary. This feature would be particularly useful when there are discrepancies between, for example, copywriting and marketing, ensuring alignment without needing to redo everything.}''

\subsubsection{Improve Efficiency to Support Design Ideation}  ($N=8$)
Participants consistently highlighted the role of AI in enhancing efficiency, allowing designers to focus more on divergent ``thinking,'' than ``doing'' repetitive tasks. For instance, P6 explained, ``\textit{I think the best thing is to accelerate your design process, like doing some repetitive thing that you had to do with some design parts. [It would be great] if AI could streamline that process for you and also edit things that we sometimes overlook.}'' 
Similarly, participants highlighted specific areas where AI could automate repetitive tasks. For example, P10 mentioned, ``\textit{Naming frames automatically would be extremely helpful, especially considering the endless number of frames we deal with in design.}'' P11 also stated, ``\textit{I can request different orientations, like landscape mode, or versions for an iPad or tablet. This automation across different screen sizes would save a ton of time.}'' 

The direct integration of AI functionalities into existing design tools was another important aspect. Participants expressed a strong desire for AI to be embedded within the tools they already use, such as Figma. For example, P19 highlighted this by saying, ``\textit{When I need textual content, I have to open ChatGPT separately. Integrating this feature directly into Figma would make it much easier for designers and save time, as there would be no need to switch between apps.}'' P12 also wanted to see AI tools integrated into their existing physical tools, stating, ``\textit{Something that would be really, really interesting would be if you could take a picture of your post-it notes, and it automatically transcribes the ideas. Maybe it recognizes the color of the post-it and what's written on it, and then it uploads everything to your digital system. That would be really cool.}'' Similarly, P9 discussed the value of bridging current tools through AI, stating, ``\textit{Right now, there's FigJam for wireframing and then you move to design. I like that separation, but it would be cool if there were an ability, maybe through AI, to take a sketch of the UI and generate a version based on that wireframe, matching the design system you've created.}''

\subsubsection{Going Beyond Texts and Prompts} ($N=7$)
Participants expressed a strong desire for AI tools that adapt to their needs and preferences as designers, who are mostly visual thinkers tackling complex problems. The current chat-based interface of most AI tools does not satisfy them. For example, P10 emphasized, ``\textit{It could be more visual while presenting data. [In the current tools], we type something, and it gives a lot of information. If that data is organized and presented better visually, it could be more useful.}''
Participants highlighted the importance of AI tools offering clear guidance and support to the designers, in terms of what the tools can do and how to do them. For example, P9 considered that ``\textit{It should include a help section or a comprehensive tutorial that provides an overview of everything the tool can do, so you're not just scratching the surface of its capabilities.}'' P13 also shared the frustration they faced in understanding AI tools, saying, ``\textit{It takes me a lot of energy to understand the AI tool. If this process improves, maybe I'll be satisfied.}''

Related, participants were often confused when interacting with the AI tools through prompting. For instance, P9 talked about this elaborately, ``\textit{I think you just need to be very specific with how you write your prompts because it directly affects the outcome or the output AI gives you. ... You have to be clear and precise. It may take some refining depending on what it gives you, so there's often a lot of back and forth involved.}'' Similarly, P17 shared their experience, saying, ``\textit{The AI's response quality can vary a lot depending on the prompt ... I need to keep refining my prompts, which is time-consuming.}'' P8 also noted, ``\textit{It's not an automatic magic tool. You need to learn how to use it effectively.}'' These challenges underscore that writing a good prompt does not come naturally for designers, and participants expressed a desire to have an alternative interaction mechanism. For example, P7 suggested that to make the interaction more comfortable, ``\textit{I would expect maybe suggestive prompts, maybe some examples of prompts in what it generates according to that. Some examples from other users.}'' Additionally, P5 envisioned AI tools that could frame questions rather than relying solely on user inputs, stating, ``\textit{It should ask me questions and it should frame its own answers.}'' 

\section{Discussion}
In this paper, we explored the integration of AI tools into the divergent thinking process of UI/UX design. We examined how designers utilize AI tools across various divergent thinking activities, including ideation, prototype exploration, and collaboration. Our findings highlighted both the opportunities and the limitations of AI in supporting design creativity. Below we first reflect on the unique concerns of AI in the context of divergent thinking, in relation to the overall UI/UX design process. We then summarize four key roles of AI to support UI/UX divergent thinking revealed by our results and outline implications for the design of future AI tools. Finally, we discuss the limitations of our current study.

\subsection{The Unique Perspectives of Using AI in Divergent Thinking of UI/UX Design}
Building on previous work about design ideation and idea management (e.g., \citet{inie_how_2020}), our research contributes by focusing on how AI can enhance the divergent thinking process for idea generation and creative exploration. Recent research has examined AI's potential in improving specific design tasks. For example, \citet{padmasiri_ai-driven_2023} emphasized AI's role in supporting tasks such as understanding user needs and facilitating rapid prototyping within the Design Thinking framework. Many previous studies have also portrayed AI as a tool for automating various design tasks to enhance efficiency and reduce manual effort, such as in \citet{sermuga_pandian_metamorph_2021} and \citet{takaffoli_generative_2024}. However, our research contributes to a more nuanced understanding of AI's role in divergent thinking. This is a phase that has not received as much attention in prior works~\cite{chiou_designing_2023}. Below, we discuss key concerns related to this perspective that are either not addressed or less emphasized in the existing literature.

\subsubsection{Efficiency gain is for unleashing creativity.}
Overall, our findings revealed that participants frequently rely on AI tools to automate routine tasks and enhance efficiency. Tasks like organizing design artifacts, generating placeholder contents with realistic examples, creating designs for multiple screens, and summarizing research findings were identified as areas where AI significantly reduced manual effort. These capabilities align with prior work~\cite{takaffoli_generative_2024, sermuga_pandian_metamorph_2021}, which emphasizes AI's efficiency in streamlining UI/UX workflows. However, our study extends these findings by illustrating how the automation of such tasks supports creativity by freeing designers to focus more on exploration and other creative design activities. After all, in the context of divergent thinking, the purpose of quickly generating content is not to create final artifacts, but to inspire and explore ideas. Therefore, efficiency gains alone are not beneficial unless they foster creativity. Overemphasizing automation and efficiency could also lead to solutions that shortcut exploration, ultimately stifling creativity. Thus, future research on AI tools for supporting divergent thinking should carefully balance automation and exploration support to fully harness the potential of this technology.

\subsubsection{Copyright and ownership may not be major concerns.}
Studies such as \citet{li_user_2024} and \citet{inie_designing_2023} highlighted prominent challenges related to copyright, ownership, bias, and data privacy in AI tools for UI/UX design support. An unexpected divergence from prior work was the absence of these concerns among our participants when discussing divergent thinking. We did not prompt this topic intentionally, and the participants also did not raise these concerns spontaneously. An explanation for this is that seeking, combining, and transforming existing design examples is a fundamental process in creative UI/UX design~\cite{Herring2009CHI, Mozaffari2022}. Even without AI, practitioners frequently get ideas from inspiration-support platforms such as Dribbble\footnote{https://dribbble.com} and Behance\footnote{https://www.behance.net}, as well as analyzing their competitors. Getting inspired by AI-generated content is not very different from this existing practice and seems natural to our participants. It is important to point out that the absence of ethical concerns is aligned with the intention of the UI/UX designers for using the AI-generated content during divergent thinking: not to reuse the material directly, but to spark new ideas. Albeit, there may be overlooked ethical issues, such as using commercial products to train AI models and the environmental impact of those models, that UI/UX designers might not consider when discussing divergent thinking. Future work is needed to explore how these concerns may impact the designers' practices.

\subsubsection{The lo-fi/hi-fi dichotomy of prototyping is becoming blurry.}
\label{sec:discussion_hifilofi}
A key tension identified in our findings is the contrast between AI-generated high-fidelity outputs and designers' traditional practice using low-fidelity methods during divergent thinking. While studies such as \citet{sermuga_pandian_metamorph_2021} and \citet{takaffoli_generative_2024} emphasized the value of AI tools in producing polished outputs quickly, our participants expressed that such tools often constrain creativity when used too early in the design process. This nuanced insight highlights a gap in existing studies on AI-based UI/UX tools, which tend to prioritize refinement over the unstructured, exploratory nature of early-stage design. At the same time, however, the use of generative AI tools tends to break the traditional dichotomy of hi-fi/lo-fi prototyping. Low-fidelity approaches, such as paper sketches and wireframes, were preferred for their flexibility and low cost to support rapid iteration during idea exploration. With the increased AI ability to quickly generate realistic prototypes, some benefits of using low-fidelity prototypes (e.g., low cost) may diminish. This is why some participants intentionally used AI to increase prototype fidelity for more accurate feedback gathering. The impacts of AI tools on prototype exploration and the iterative design practice require further investigation in future work.

\subsection{Facilitating the Four Key Roles of AI to Support Divergent Thinking in UI/UX Design}
Our results on UI/UX designers' current divergent thinking practices and their use of AI tools in this process indicated four roles AI can play in supporting divergent thinking in UI/UX design. Although important for divergent thinking, these roles are less emphasized in the literature about AI support in the general UI/UX design process. In the following sections, we discuss these four roles, their connections with the literature, and opportunities for designing AI tools to support each role.

\subsubsection{Gathering and synthesizing research data}
Our participants considered research and discovery, such as user research and competitive analysis, crucial activities for divergent thinking in UI/UX design. They discussed ways in which AI could help with completing these activities, ranging from gathering information about industrial trends, synthesizing and summarizing research data, and creating pre-design artifacts like personas and flowcharts. This type of AI uses echoes findings by~\citet{takaffoli_generative_2024}, who identified that UI/UX professionals reported more comprehensive use of GenAI in research-related activities than in design-related activities. However, a recent literature review on collaboration between designers and AI~\cite{shi_understanding_2023} found that only very few previous studies have focused on this aspect of the UI/UX process. Although these activities may constitute a small part of the overall UI/UX process, they are crucial for driving divergent thinking and play a vital role in shaping innovative design solutions. Our results highlighted the importance of this gap and called for more future research to explore this area.

A delicate concern related to this role of AI is to achieve efficiency while preserving the ``reflection-in-action'' of UI/UX practitioners~\cite{Schon1984Reflective}. Overly automating the data gathering and analyzing process may stifle critical reflection, eventually hindering practitioners' creativity during divergent thinking. This is related to the tension we identified between automation and creative control. Thus, when supporting research and data analysis, future AI tools should focus on fostering reflection. For example, instead of providing one summarization of the user research results, tools could consider providing multiple summaries from different perspectives and let UI/UX designers judge their relevance and reflect on ways to adjust research strategies. Moreover, AI's ability to provide real-time support during user study sessions (e.g., suggesting follow-up questions based on user inputs) will further encourage reflection and streamline the user research process.

\subsubsection{Kick-starting the creative design process} 
During the divergent thinking process, designers may struggle with mental blocks or a fear of failure, which can prevent them from starting to produce creative solutions. Our participants highlighted how AI tools can help alleviate these challenges by providing a starting point for innovation. Specifically, this role of AI was mentioned both at the very beginning of the divergent thinking process, acting as a catalyst to inspire and stimulate general ideas, and during the prototyping phase, to generate preliminary solutions for further adjustments and explorations. This role of AI tools is unique to divergent thinking and is rarely elaborated in related work focusing on AI for the general UI/UX design process.

Notably, a traditional way of kick-starting the creative design process is through low-fidelity methods such as sketching~\cite{Buxton2007Sketching}. According to our participants, these methods remain a cornerstone of divergent thinking due to their flexibility and adaptability. While focusing on kick-starting the creative process, future AI tools should complement rather than replace these traditional approaches. For example, AI could assist by recommending initial design concepts or offering suggestions based on research data, while still leaving room for designers to refine and iterate through hands-on, low-fidelity techniques like sketching. Keeping the generated prototypes somehow ``low-fidelity'' in the beginning (e.g., by reducing the resolution or presenting them in the style of sketches or wireframes) could help stimulate creative exploration and avoid overstepping early-stage flexibility.

\subsubsection{Creating alternatives for inspiration}
Ideation and exploring design options are considered crucial activities of divergent thinking in UI/UX design. Once having some initial ideas, participants frequently discussed the use of AI tools to create alternatives for inspiration. This type of usage was mentioned by our participants to support the exploration of various design aspects (e.g., layout and color selection) and particular UI components. This role of AI in divergent thinking is related to addressing \textit{design fixation}, ``a blind adherence to a set of ideas or concepts''~\cite{JANSSON1991Fixation}. Researchers have previously explored the use of AI in addressing design fixation. However, those efforts are either not dedicated to UI/UX design (e.g., graphic design~\cite{wadinambiarachchi_effects_2024} and fashion design~\cite{Jeon2021}) or only provide limited user control (e.g.,~\cite{Mozaffari2022}).

One of the central values of designers is the need to preserve their creative control while leveraging AI's capabilities. When generating design alternatives, future AI tools should enable flexibility in making precise adjustments (either manual changes or regenerating certain parts of the UI for further exploration) without compromising the overall creative vision. Tools should also support the iterative design exploration process, allowing falling back and preserving multiple versions of design alternatives (e.g., through version control). Moreover, our participants frequently reported that while AI tools are helpful in generating a large volume of ideas, these ideas often fail to meet the specific contextual requirements of their projects whether related to branding, market, or user personas. To address this, AI tools must offer more contextually aware suggestions that incorporate the specific parameters of a project.

\subsubsection{Altering prototype fidelity for feedback gathering}
Our participants appreciated the ability of AI to generate high-fidelity prototypes and UI components. This benefit of AI in supporting UI/UX design is indeed widely discussed in the literature~\cite{shi_understanding_2023, inie_designing_2023}. However, when focusing on divergent thinking, our participants emphasized that the increased fidelity that AI brings to the table is particularly beneficial for gathering accurate feedback from other stakeholders. They used AI to generate textual content, visual images, and UI components with a particular style to create prototypes that look closer to the end product. This level of realism would help stakeholders better visualize the design, leading to more actionable insights and informed decision-making early in the design process.

However, caution should be taken when using AI to generate prototypes with unnecessarily high fidelity, as this may introduce details that distract feedback from the main focus of the current design stage. This is related to the potential of AI to break the hi-fi/lo-fi dichotomy in UI/UX design (see Section~\ref{sec:discussion_hifilofi}). Future AI tools for divergent thinking should consider enabling designers to adjust the fidelity level of the generated content. The level of fidelity could also be automatically adapted to the current design stage and/or the overall context of the prototype. Moreover, AI tools could also improve the feedback loop by filtering the irrelevant feedback, automating the organization and synthesis of feedback, and even suggesting directions for further explorations.

\subsection{Limitations and Future Work}
Our study provides valuable insights into the integration of AI tools in supporting divergent thinking in UI/UX design, but several limitations must be acknowledged. Although the participant pool included designers with a range of experiences, its geographic focus was mainly on North America and Asia (Canada, Mexico, Iran, India, and Pakistan) because of the geographic location and social connection of the authors. Design practices and perceptions towards AI can vary across regions and cultures, and the results may not reflect global perspectives, which can be investigated in future work.
Furthermore, the focus on designers who have industrial experiences leaves out the perspectives of novice designers and students, whose perceptions and interactions with AI tools may differ. This presents an opportunity for future research to explore how AI tools support or hinder creativity and learning for less experienced designers in the divergent thinking phase of the design process.
Finally, we are very well aware that AI technologies evolve rapidly and the findings of this study may not capture future developments. Questions remain regarding how AI tools will be further integrated into design workflows as they rapidly advance. Continuous efforts tracking the evolving role of AI in design over time could ensure that research remains current with technological advancements.

\section{Conclusion}
In this study, we investigated how UI/UX designers perceived the role of AI tools in the divergent thinking phases of the UI/UX design process. We found that, in this context, designers valued AI tools that offer greater control over the ideation process, streamline collaboration and feedback loops, automate repetitive tasks to liberate creativity, and go beyond textual prompting. Our results indicated unique concerns related to AI for divergent thinking that can be addressed in future work, including balancing efficiency and exploration, examining the distinct ethical impacts, and investigating emerging design ideation paradigms with AI. Moreover, we revealed promising roles of AI in supporting divergent thinking, through facilitating UI/UX-related research, kick-starting the creative process, generating design alternatives, and facilitating prototype exploration. By reflecting on our results, we provided specific design opportunities to enhance these roles. Our work contributes to advancing our understanding of how AI tools can be better integrated to foster more innovative UI/UX design practices.

\begin{acks}
We thank our participants for their time and valuable insights. We also thank the anonymous reviewers for helping us improve the paper. This work is partially supported by the Canada Research Chairs program (CRC-2021-00076) and the Natural Sciences and Engineering Research Council of Canada (RGPIN-2018-04470).
\end{acks}

\balance

\bibliographystyle{ACM-Reference-Format}
\bibliography{references}

\end{document}